%% file: main.tex
\documentclass[pdflatex,sn-nature]{sn-jnl}
\usepackage{graphicx}%
\usepackage{multirow}%
\usepackage{amsmath,amssymb,amsfonts}%
\usepackage{amsthm}%
\usepackage{mathrsfs}%
\usepackage[title]{appendix}%
\usepackage{xcolor}%
\usepackage{textcomp}%
\usepackage{manyfoot}%
\usepackage{booktabs}%
\usepackage{algorithm}%
\usepackage{algorithmicx}%
\usepackage{algpseudocode}%
\usepackage{listings}%
\usepackage{bm}
\input{header}
\theoremstyle{thmstyleone}%
\theoremstyle{thmstyletwo}%

\theoremstyle{thmstylethree}%

\begin{document}

\title[Magnetized ISM turbulence]{The spectrum of magnetized turbulence in the interstellar medium}


\author*[1,2,3]{\fnm{James} \sur{R. Beattie}}\email{james.beattie@princeton.edu}
\author[3,4]{\fnm{Christoph} \sur{Federrath}}
\author[5,6,7,8]{\fnm{Ralf S.} \sur{Klessen}}
\author[9]{\fnm{Salvatore} \sur{Cielo}}
\author[1]{\fnm{Amitava} \sur{Bhattacharjee}}

\affil[1]{\orgdiv{Department of Astrophysical Sciences}, \orgname{Princeton University}, \orgaddress{\city{Princeton}, \postcode{08540}, \state{NJ}, \country{USA}}}

\affil[2]{\orgdiv{Canadian Institute for Theoretical Astrophysics}, \orgname{University of Toronto}, \orgaddress{\city{Toronto}, \postcode{M5S3H8}, \state{ON}, \country{Canada}}}

\affil[3]{\orgdiv{Research School of Astronomy and Astrophysics}, \orgname{Australian National University}, \orgaddress{\city{Canberra}, \postcode{2611}, \state{ACT}, \country{Australia}}}
 
\affil[4]{\orgname{Australian Research Council Center of Excellence in All Sky Astrophysics (ASTRO3D)}, \orgaddress{\city{Canberra}, \postcode{2611}, \state{ACT}, \country{Australia}}} 

\affil[5]{\orgdiv{Zentrum f\"ur Astronomie, Institut f\"ur Theoretische Astrophysik}, \orgname{Universit\"at Heidelberg}, \orgaddress{\city{Heidelberg}, \postcode{69120}, \state{Baden-W\"urttemberg}, \country{Germany}}} 

\affil[6]{\orgdiv{Interdisziplin\"ares Zentrum f\"ur Wissenschaftliches Rechnen}, \orgname{Universit\"at Heidelberg}, \orgaddress{\city{Heidelberg}, \postcode{69120}, \state{Baden-W\"urttemberg}, \country{Germany}}} 

\affil[7]{\orgdiv{Center for Astrophysics}, \orgname{Harvard \& Smithsonian}, \orgaddress{\city{Cambridge}, \postcode{02138}, \state{MA}, \country{USA}}}

\affil[8]{\orgdiv{Elizabeth S. and Richard M. Cashin Fellow}, \orgname{Radcliffe Institute for Advanced Studies at Harvard University}, \orgaddress{\city{Cambridge}, \postcode{02138}, \state{MA}, \country{USA}}}

\affil[9]{\orgname{Leibniz Supercomputing Center of the Bavarian Academy of Sciences and Humanities}, \orgaddress{\city{Garching}, \postcode{85748}, \state{Bavaria}, \country{Germany}}}  


\abstract{\textbf{The interstellar medium (ISM) of our Galaxy is magnetized, compressible and turbulent, influencing many key ISM properties, like star formation, cosmic ray transport, and metal and phase mixing. Yet, basic statistics describing compressible, magnetized turbulence remain uncertain. Utilizing grid resolutions up to $10,\!080^3$ cells, we simulate highly-compressible, magnetized ISM-style turbulence with a magnetic field maintained by a small-scale dynamo. We measure two coexisting kinetic energy cascades, $\ekin(k) \propto k^{-n}$, in the turbulence, separating the plasma into scales that are non-locally interacting, supersonic and weakly magnetized $(n=2.01\pm 0.03\approx 2)$ and locally interacting, subsonic and highly magnetized $(n=1.465\pm 0.002\approx 3/2)$, where $k$ is the wavenumber. We show that the $3/2$ spectrum can be explained with scale-dependent kinetic energy fluxes and velocity-magnetic field alignment. On the highly magnetized modes, the magnetic energy spectrum forms a local cascade $(n=1.798\pm 0.001\approx 9/5)$, deviating from any known \textit{ab initio} theory. With a new generation of radio telescopes coming online, these results provide a means to directly test if the ISM in our Galaxy is maintained by the compressible turbulent motions from within it.}}%

\maketitle
\newpage
\section{Introduction}\label{sec:intro}
	In the interstellar medium (ISM) of our Galaxy, the coupling between turbulence and the magnetic fields plays an important, multifaceted role. In the cold ($T \approx10\;\rm{K}$) molecular phase of the ISM, it changes the ionization state of the plasma by controlling the diffusion of cosmic rays \citep{Krumholz2020_cosmic_ray_transport_SB,Xu_2022_damping_in_ISM,Beattie2022_ion_alfven_fluctuations,Sampson2023_SCR_diffusion,Ruszkowski2023_CR_review}, gives rise to the filamentary structures that shape and structure the initial conditions for star formation \citep{Federrath2016_filaments,Hacar2023_filaments_ICs_for_SF}, and through turbulent and magnetic support, changes the rate at which the cold plasma converts mass density into stars \citep{MacLow2004,McKee2007,Hennebelle2012,Federrath2012,Burkhart2018,Nam2021}. In the plasma rest frame, the root-mean-squared (rms) turbulent velocity fluctuations $u_0 \approx 2\;\rm{km}\rm{s}^{-1}$ on the outer scale of the cold plasma $\ell_0 \approx 10\;\rm{pc}$ are supersonic $\M =u_0/c_s \approx 10$, where $c_s \approx 0.2 \; \rm{km}\rm{s}^{-1}$ is the sound speed and $\M$ is the turbulent sonic Mach number. Furthermore, on these scales, the plasma Reynolds number $\Re = u_0 \ell_0 / \nu \sim 10^6-10^9$ is large, where $\nu$ is the coefficient of kinematic viscosity \citep{Beattie2019b,Krumholz2015_star_formation_text,Federrath2016_brick,Ferriere2020_reynolds_numbers_for_ism}. $\Re$ determines the range of scales that are within the turbulent cascade, $\ell_\nu < \ell < \ell_0$, where $\ell_\nu$ is the viscous dissipation scale. Under the assumptions of incompressibility, homogeneity and isotropy, and constant energy flux $\varepsilon = u_0^3 / \ell_0$ between neighboring turbulent modes, Kolmogorov \citep{Kolmogorov1941} predicts $\ell_\nu \sim \Re^{-3/4}\ell_0$; hence, to measure cascade physics, one has to simulate the turbulence with as many resolved scales as possible, on the largest grids possible. 
    
    Hydrodynamical ISM-type turbulence has been simulated at $\Re \gtrsim 10^6$, approaching realistic ISM $\Re$ on a $10,048^3$ grid, providing enough dynamical range to demonstrate the existence of two scale-separated power laws in the kinetic energy spectra \citep{Ferrand2020_10k3_fluxes,Federrath2021}, as opposed to a single Kolmogorov-type power law. Based on the second-order structure functions, scales $\ell_s < \ell < \ell_0$ exhibit a Burgers spectrum, with $\ekin(k) \sim k^{-2}$ \citep{Burgers1948}, while scales $\ell_\nu < \ell < \ell_s$ follow a Kolmogorov spectrum, with $\ekin(k) \sim k^{-5/3}$ (with intermittency corrections; \citep{She1994}), where $k = 2\pi/\ell$ is the wavenumber, $\ekin(k)$ is the kinetic energy spectrum and $\ell_s \sim \ell_0/\M^{2} $ is the sonic scale, where $\delta u(\ell_s) = c_s$ is a critical scale for turbulence-regulated star formation theories \citep{Krumholz2005,Federrath2012}, where $\delta u(\ell)$ is the velocity dispersion on scale $\ell$. No such simulation exists for supersonic, magnetized turbulence at $\Re \gtrsim 10^6$, and it is unknown how an additional magnetic field will modify the Burgers or Kolmogorov spectra, nor how $\ell_s \sim \ks^{-1}$ responds to the additional magnetic fluctuations. 
    
    The ISM plasma is permeated by a dynamically significant magnetic field that can be in energy equipartition with the hydrodynamic turbulence \citep{Federrath2016_brick,Hu2019,Skalidis2021_obs_sub_alf}. The field restructures the ISM, creating a network of organized mass density and magnetic structures \citep{Gaesnsler_2011_trans_ISM,Soler2017,Clark2019}. The strong magnetic fields are inevitably maintained through the generation of magnetic energy through a turbulent dynamo \citep{Schekochihin2004_dynamo,Rincon2019_dynamo_theories}. The additional magnetic fluctuations significantly alter the physics of the turbulence cascade via shear Alfv\'en mode interactions \citep{Goldreich1995,Boldyrev2006} and magnetic and velocity correlations \citep{Boldyrev2006,Mason2006_dynamical_alignment,Banerjee2023_relaxed_states}.  
    
    Similar to supersonic turbulence, recent numerical evidence has been accumulating that suggests that at high enough magnetic Reynolds number $\Rm = u_0 \ell_0 / \eta$, where $\eta$ is the plasma resistivity, there may be a change in the cascade of three-dimensional MHD turbulence \citep{Dong2022_reconnection_mediated_cascade,Galishnikova2022_saturation_and_tearing,Fielding2022_ISM_plasmoids}. The large-$\Rm$ theories combine magnetic reconnection and turbulence together, where reconnection-driven tearing instabilities disrupt and modify a cascade of sheet-like eddies. They predict $\emag(\kperp) \sim \ekin(\kperp) \sim \kperp^{-11/5}$, where $\kperp$ is the wavevector perpendicular to the large-scale magnetic field \citep{Mallet2017_plasmoid_disruptions, Comisso2018_MHD_turbulence_plasmoid_regime,Boldyrev2020_tearing_mode_instability,Dong2022_reconnection_mediated_cascade}. To separate the cascade timescales from instability timescales, simulations require extremely high resolutions that support $\Rm\gtrsim 10^5$, ensuring that thin current sheets (formed from anisotropic eddies), found ubiquitously in global and local ISM simulations \citep{Fielding2022_ISM_plasmoids,Ntormousi2024_strong_turbulence_current_sheets_galaxies}, become unstable to the plasmoid instability that yields a nonlinear regime of fast magnetic reconnection \citep{Bhattacharjee2009_fast_reconnection,Uzdensky2010_fast_reconnection}. This happens on scales where the instability growth timescale is shorter than the cascade timescale of the turbulence, leading to a break scale $k_*^{-1} \sim \ell_{*} \sim \Rm^{-4/7}$ in the cascade \citep{Loureiro2017_reconnection_in_turbulence,Dong2022_reconnection_mediated_cascade}. The scale, $\ell_{*}$, has been measured in decaying MHD turbulence at $\Rm \sim 10^5$ \citep{Dong2022_reconnection_mediated_cascade}, as well as there being observed signatures of the tearing instability in local, multiphase ISM simulations \citep{Fielding2022_ISM_plasmoids}. Furthermore, there is tentative evidence of a tearing-mediated cascade $\ekin(k) \sim k^{-19/9}$ in the saturated state of the subsonic turbulent dynamo \citep{Galishnikova2022_saturation_and_tearing}. 
    
    Most of the theories and extremely high-resolution models that capture the multi-scale nature of magnetized turbulence are for subsonic and incompressible plasmas, often with a uniform background magnetic field \citep{Iroshnikov_1965_IK_turb,Kraichnan1965_IKturb,Goldreich1995,Boldyrev2006,Beresnyak2014_4k_incomp_sim,Dong2022_reconnection_mediated_cascade}, potentially limiting their applicability for understanding the basic statistical properties of the compressible ISM turbulence supported by a dynamo. Indeed, it is an important and open question to understand the fundamentals of the cascade, including the spectra, the energy flux, and the hierarchy of scales in highly compressible MHD turbulence regimes that are prevalent in our Galaxy.

    \begin{figure}
        \centering
        \includegraphics[width=\linewidth]{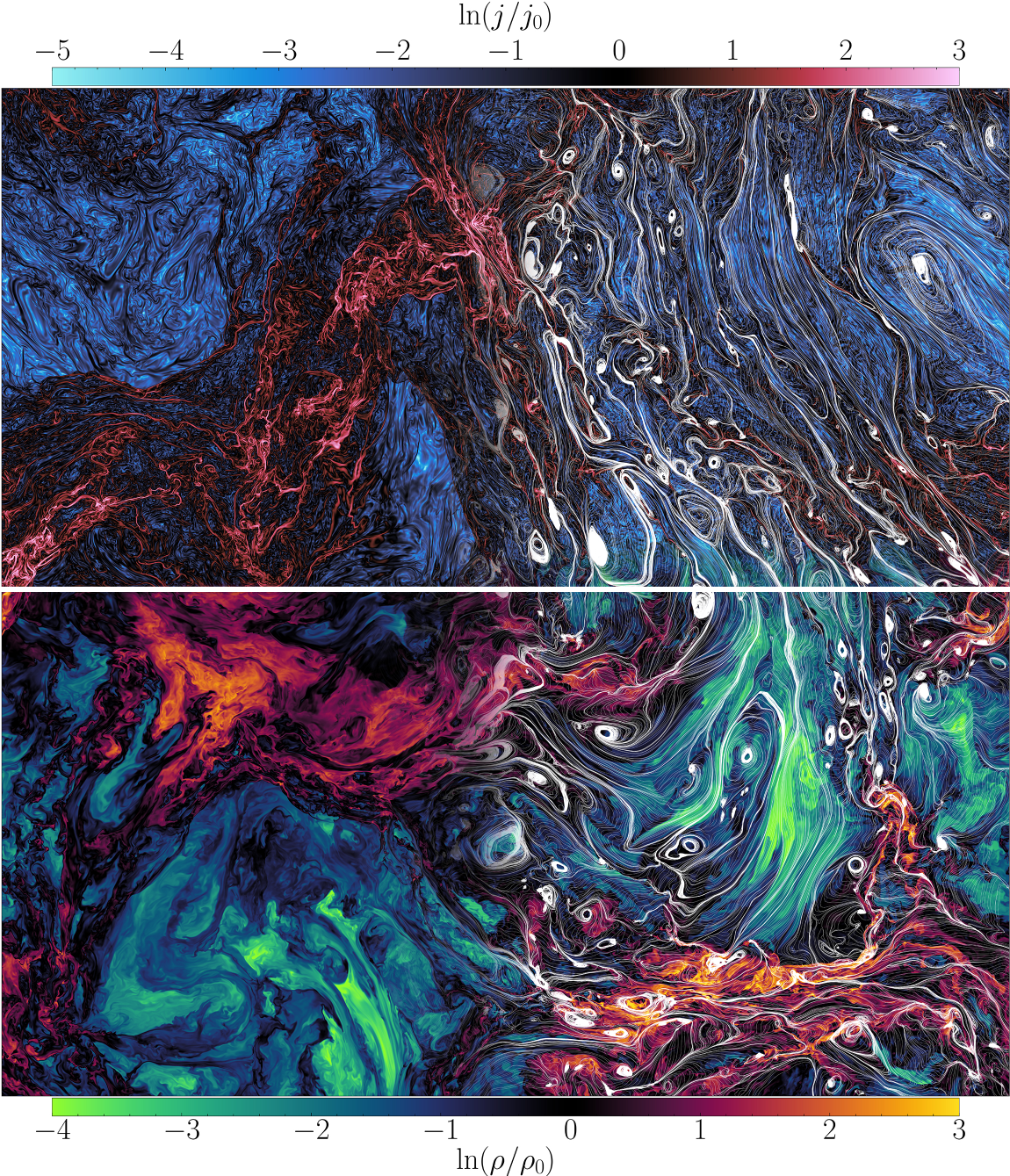}
        \caption{\textbf{The current density, magnetic field and mass density structure in the world's largest supersonic MHD turbulence simulation.} A two-dimensional slice of the $10,\!080^3$ simulation showing the logarithmic magnitude of the current density $\ln(j/j_0)$ (top), mass density $\ln(\rho/\rho_0)$ (bottom), and magnetic field lines sliced on the plane (right; white), where the subscript zero indicates the volume average. Fractal current sheets, shown in red in the top-left of the figure, are strongly correlated with the densest regions, shown in yellow in the bottom-left of the figure, accompanied by tightly coiled magnetic fields.}
        \label{fig:slice_poster}
    \end{figure}
    
\section{Results}\label{sec:results}
    \subsection{Supersonic MHD turbulence at unprecedented resolutions}
        We present the first results from an ensemble of driven, supersonic, $\M  =  4.3 \pm 0.2$, magnetized turbulence simulations with a magnetic field that is self-consistently maintained by the turbulent, small-scale dynamo in saturation, providing a volume integral energy ratio of $\emag/\ekin = 0.242 \pm 0.022$ (see Appendix~\ref{app:volume_integral_quants} for more details). Dynamo-generated magnetic fields have been shown to better reconstruct the observational $B-\rho$ relation compared to imposed large-scale field simulations \cite{Whitworth2024_magnetic_gas_relation}. The grids vary from $2,\!520^3$ ($\Re \sim \Rm \sim 10^5$) up to $10,\!080^3$ ($\Re \sim \Rm \gtrsim 10^6$; see Appendix~\ref{app:reynolds_numbers} for details), approaching the $\Re$ of the cold phase ISM and larger than the $\Re$ in the warmer phases \citep{Ferriere2020_reynolds_numbers_for_ism}, meaning that the scale-separation between the inner and outer turbulent scales is realistic for the ISM. The simulations are discretized on a triply-periodic domain with length $L$. In Figure~\ref{fig:slice_poster}, we visualize a two-dimensional slice of the logarithmic current and mass density, $\ln(j/j_0)$ and $\ln(\rho/\rho_0)$, respectively, with magnetic field streamlines shown in white. The zero subscript indicates the mean over the entire volume. The $\ln(j/j_0)$ field shows fractal $\ln(j/j_0) > 0$ current sheet structures in red, and $\ln(j/j_0) < 0$ current voids in blue, whilst the $\ln(\rho/\rho_0)$ field shows shocked, high-density $\ln(\rho/\rho_0) > 0$ filaments in red, and deep $\ln(\rho/\rho_0) < 0$ voids in green, with fluctuations $-4 \leq \ln(\rho/\rho_0) \leq 3$, highlighting how the mass and current density vary by many orders of magnitude.
        
        Presently, these are the largest supersonic, magnetized simulations in the world, almost an order of magnitude larger in grid resolution compared to previous simulations in this regime \citep{Grete2023_as_a_matter_of_dynamical_range}, and are the first MHD simulations to resolve both a supersonic and subsonic cascade with a self-consistent dynamo-sustained magnetic field. The simulations utilized over 80~million CPU hours distributed across nearly $140,\!000$~compute cores on the high-performance supercomputer, SuperMUC-NG, at the Leibniz Supercomputing Centre. We integrate the $10,\!080^3$ simulation for $t\approx 2t_0$, where $t_0 = \ell_0 / u_0$ is the turnover time on the driving scale of the turbulence $\ell_0 = L/2$ (or equivalently $k_0L/2\pi = 2$), allowing for time-averaging of all key statistics across $t \approx 2t_0$, making for robust, statistically significant results. We provide details on the simulation methods in Appendix~\ref{app:simulations}.
        
        \begin{figure}
            \centering
            \includegraphics[width=0.95\linewidth]{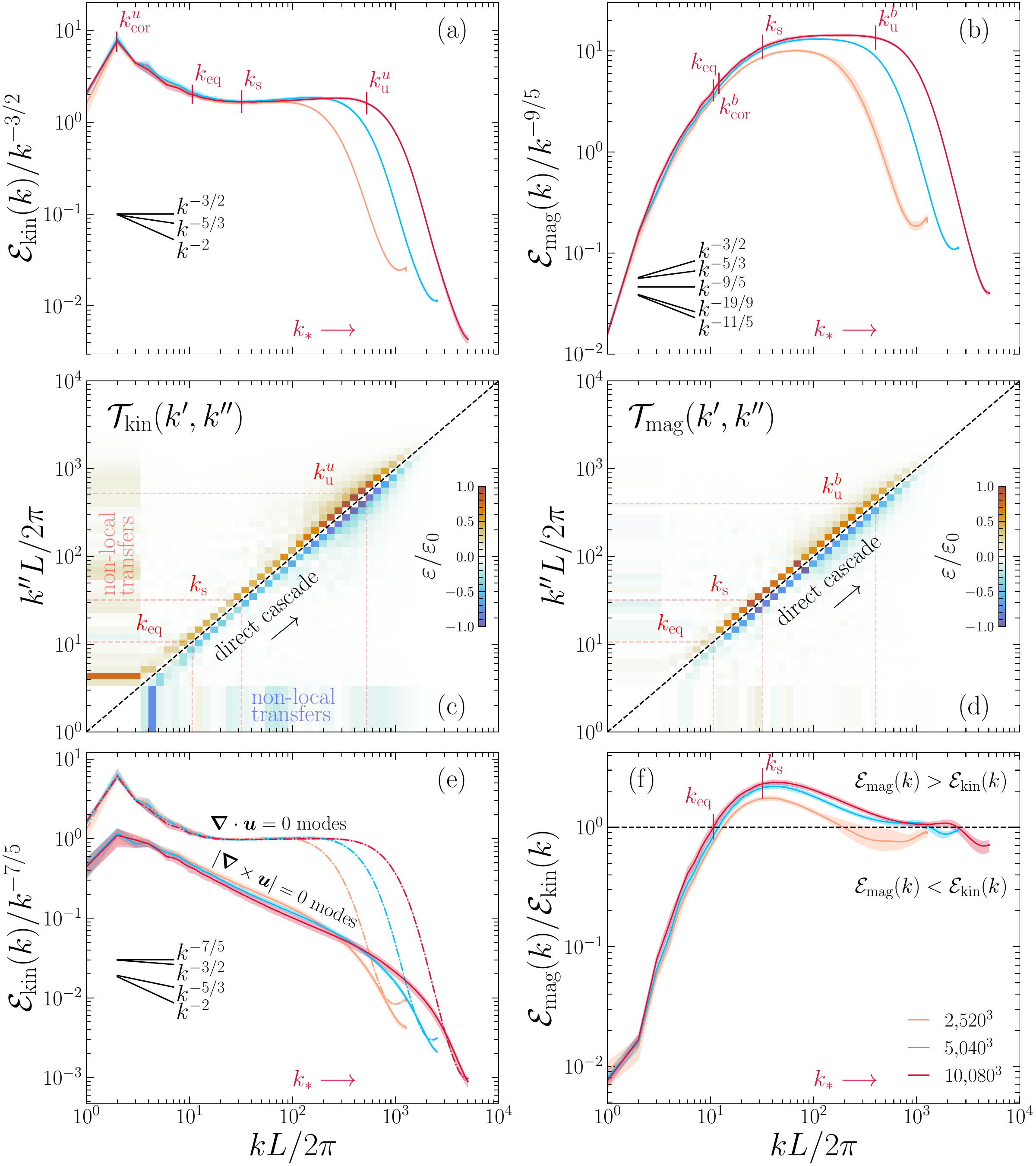}
            \caption{\textbf{The energy spectra, fundamental turbulence scales and energy flux transfer functions.} The correlation scale, $\kcor$, energy equipartition scale, $\keq$, sonic scale, $\ks$, inner scale, $\ku$, and upper bound for the plasmoid outer scale, $k_{*}$, are annotated in the panels (see Appendix~\ref{app:turbulent_scales} for definitions). Each color corresponds to a different grid resolution. \textbf{(a)}: The kinetic energy spectra $\ekin(k)$ compensated by $k^{-3/2}$. We observe a transition between scales separated by $\keq$, where the turbulence goes from hydrodynamically to magnetically dominated (see panel~d). On large scales, $\ekin(k) \sim k^{-2}$, whilst on small scales, $\ekin(k) \sim k^{-3/2}$. \textbf{(b)}: The magnetic energy spectrum $\emag(k)$ compensated by $k^{-9/5}$, with the same annotated scales as in (a). The compensation shows an extended power law, $\emag(k) \sim k^{-9/5}$, at $k > \ks$. Different turbulence power-law models are shown for comparison: dynamical alignment, $k^{-3/2}$ \citep{Boldyrev2006}, unaligned strong turbulence, $k^{-5/3}$ \citep{Goldreich1995}, an empirical relation, $k^{-9/5}$ \citep{Fielding2022_ISM_plasmoids}, tearing instability in dynamo, $k^{-19/9}$ \citep{Galishnikova2022_saturation_and_tearing}, and the reconnection-mediated cascade, $k^{-11/5}$ \citep{Dong2022_reconnection_mediated_cascade}. \textbf{(c)}: the kinetic energy transfer function for both kinetic energy cascade terms, showing strong non-local transfers of flux from the largest scales, coupled with a direct, local cascade on $k\lesssim \ku^u$. \textbf{(d)}: the same as (c) but for the magnetic energy, showing the formation of a cascade on $\keq \lesssim k \lesssim \ku^b$. \textbf{(e)}: The $\ekin(k)$ decomposed into compressible ($|\bm{\nabla}\times\bfu|=0$; solid line) and incompressible ($|\bm{\nabla}\cdot\bfu|=0$; dashed line) modes. Each spectrum traces a different slope, suggesting compressible modes are not passive to incompressible modes \citep{Lithwick2001_compressibleMHD}. \textbf{(f)}: The $\emag(k)/\ekin(k)$ energy ratio showing the $\keq$ transition, and the peak of $\emag(k)/\ekin(k)$ at $\sim\ks$.}
            \label{fig:spectra}
        \end{figure}

        \begin{figure}
            \centering
            \includegraphics[width=\linewidth]{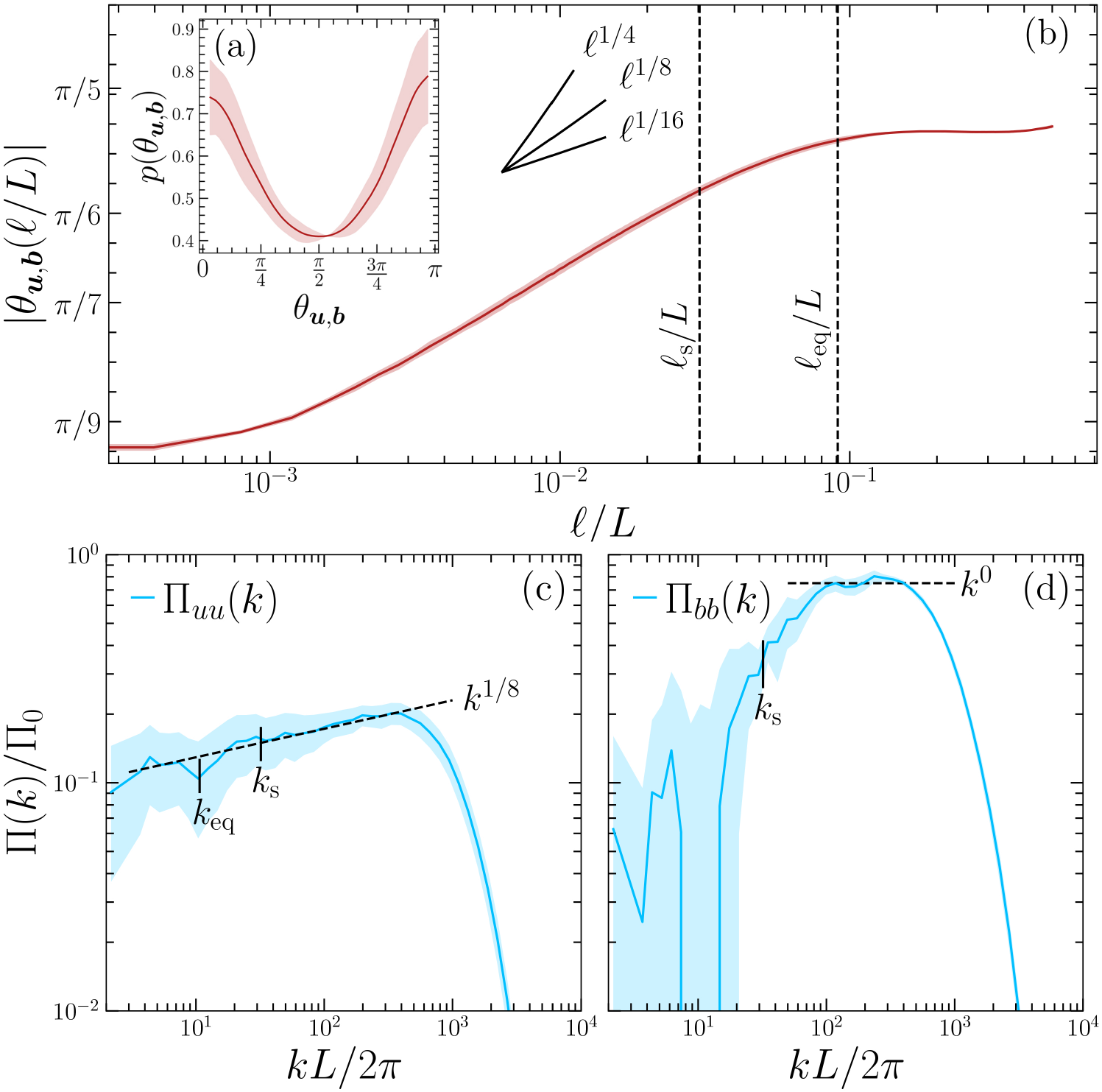}
            \caption{\textbf{The scale-dependent alignment of the velocity and magnetic field and energy flux.} \textbf{(a)}: the global probability distribution function of $\theta_{\bfu,\bfb}$. \textbf{(b):} the scale-dependent absolute local angle $|\theta_{\bfu,\bfb}|$ between $\bfu$ and $\bfb$. The angle $|\theta_{\bfu,\bfb}| \sim \ell^{1/8}$ shows scale-dependent alignment at scales smaller than the energy equipartition scale $\ell_{\rm eq}$, indicating that the nonlinearities in the turbulence become progressively weaker throughout the subsonic cascade. The relation, $|\theta_{\bfu,\bfb}| \sim \ell^{1/8}$ is inconsistent with the dynamical alignment prediction $|\theta_{\bfu,\bfb}| \sim \ell^{1/4}$ \citep{Boldyrev2006,Perez2009_dynamical_alignment_of_imbalanced_islands}, and is currently unexplained by any scale-dependent alignment theory. \textbf{(c):} the cross-scale kinetic energy flux, $\Pi_{uu}(k)$, which depends upon scale, $\Pi_{uu}(k) \sim k^{1/8}$ (shown by the dashed line). \textbf{(d):} the cross-scale magnetic energy flux, $\Pi_{bb}(k)$, which is approximately constant within the $\emag(k) \sim k^{-9/5}$ cascade, $\Pi_{bb}(k) \sim k^{0}$. }
            \label{fig:alignment_between_ub}
        \end{figure}
    
    \subsection{Energy spectra, flux and fundamental scales}
        This unique numerical experiment resolves a broad range of important scales in ISM-type turbulence and profoundly challenges the fundamental tenets of MHD turbulence theories. We show this directly by measuring the time-averaged, isotropic $\ekin(k)$ (a) and $\emag(k)$ (b) in Figure~\ref{fig:spectra} (see Appendix~\ref{app:turbulent_scales} for spectra and scale definitions) and deriving a number of important scales directly for them. We further report the energy flux transfer functions, $\mathcal{T}(k',k'')$ between pairs of $k$ mode shells, $k'$ and $k''$. (see Appendix~\ref{app:transfer_functions} for definitions; \citep{Grete2023_as_a_matter_of_dynamical_range}). We report all empirical measurements of slopes for each spectrum in Appendix~\ref{app:slopes}.
        
        As has been observed for the kinetic energy spectrum of supersonic hydrodynamical turbulence, in Figure~\ref{fig:spectra}~(a) we find a Burgers spectrum $\ekin(k) \sim k^{-2}$ on large scales \citep{Burgers1948}, where $\M > 1$, $\kcor^u = (2.03 \pm 0.01) 2\pi/L \approx k_0 > k > \keq = (10.6\pm 0.7) 2\pi/L$, with $\kcor^u$ the correlation scale of the turbulence, and $\keq$ the $\ekin(\keq) = \emag(\keq)$ energy equipartition (or MHD) scale. The kinetic energy transfer functions \citep{Grete2017_shell_models_for_CMHD}, $\mathcal{T}_{uu}(k',k'')$, Figure~\ref{fig:spectra}~(c), demonstrate that transfer is strongly non-local from these scales, reaching all the way down to the dissipation scale, shown through the off-diagonal fluxes. This is exactly what one expects from the supersonic range of scales within the turbulence \citep{Galtier2011_exact_relations,Ferrand2020_10k3_fluxes}. 
        
        On the $\M < 1$ range of scales $\ks =(32\pm2) 2\pi/L < k < \ku^u$, where $\ku^u$ is the inner scale, and $\ks$ the sonic scale, the characteristic scale of the velocity gradients, we find an Iroshnikov-Kraichnan (IK)-type spectrum $\ekin(k) \sim k^{-3/2}$ \citep{Iroshnikov_1965_IK_turb,Kraichnan1965_IKturb}. A similar break between spectral scalings has been found in $\M \approx 4$ hydrodynamical turbulence, where $\ks$ is consistent within 1$\sigma$ to the hydrodynamical $\ks$ measured in \citep{Federrath2021}, even with the additional effect of the magnetic field, showing that $\ks$ is set by the large-scale $\M > 1$ motions, and the subsonic cascade by the super-equipartition energy magnetic field. On this range of scales, there is a dominant forward energy flux locally between neighboring $k$ modes, as indicated in Figure~\ref{fig:spectra}~(c). This shows that $\ekin(k) \sim k^{-3/2}$ is a classical, local cascade.
                
        Figure~\ref{fig:spectra}~(b) demonstrates that no dichotomy exists in $\emag(k)$, which has a single self-similar range of scales on $\ks < k < \ku^b$, undergoing a direct, local cascade, shown with $\mathcal{T}_{bb}(k',k'')$ in Figure~\ref{fig:spectra}~(d), where $\ku^b$ is the magnetic inner scale, with scaling $\emag(k) \sim k^{-9/5}$, which is unexplained by any current MHD turbulence theory, including theories for Alfv\'enic turbulence \citep{Goldreich1995}, scale-dependent Alfv\'enic alignment \citep{Boldyrev2006}, and high-$\Rm$ tearing instabilities  \citep{Galishnikova2022_saturation_and_tearing,Dong2022_reconnection_mediated_cascade}. Similar $\emag(k)$ has been previously measured in $\M \lesssim 1$ turbulence on grid resolutions of $\approx 2,\!000^3$ in \citep{Fielding2022_ISM_plasmoids}. Based on the approximate physical size scale of the plasmoid structures in the voids (See Appendix~\ref{app:current_structure} and Figure~\ref{fig:plasmoids}), we show the upper bound of the plasmoid scale, $k_{*}$ on each of the panels. We see no significant spectral steepening on or below these scales, most likely due to the low volume-filling factor of the unstable sheets that we described in the previous section. The magnetic correlation scale, $\kcor^b = (12.05 \pm 0.05)2\pi/L$ is associated with $\keq$ and not the energy injection scale $k_0$, meaning the magnetic field is correlated on significantly smaller scales $\kcor^{b}/\kcor^{u} \approx 6$ compared to the velocity. In Figure~\ref{fig:spectra}~(f), we show that the $\emag(k)/\ekin(k)$ spectrum is maximized at $\ks$, hence the sonic transition $\delta u(\ks) = c_s$ is the most magnetized scale in the turbulence. 
        
        In Figure~\ref{fig:spectra}~(e) we decompose the velocity into incompressible $\bm{\nabla}\cdot\bfu = 0$, $\bfu_s$, and compressible $|\bm{\nabla}\times\bfu| = 0$, $\bfu_c$ mode spectra, which both play diverse and different roles in Galactic turbulence \citep{Beattie2025_supernova_driven_turbulence}. The $\bfu_c$ spectrum follows roughly $\ekin(k) \sim k^{-2}$ for all $k$, and the $\bfu_s$ spectrum closely matches the total kinetic spectrum, with a slightly shallower spectrum $\ekin(k) \sim k^{-7/5}$ on the $k > \keq$ scales. This shows that $\bfu_c$ modes are not passively tracing the $\bfu_s$, as is the standard result from previous compressible MHD theories \citep{Lithwick2001_compressibleMHD}. Instead, they are potentially either transported non-locally via shocks across all scales, similarly to what is observed in the low-$k$ modes in panel~(c), or undergoing their own fast, weak cascade \citep{Boldyrev2009_spectrum_of_weak_turb}. Regardless, the plasma becomes incompressible on small scales due to the steep $\ekin(k) \sim k^{-2}$ spectrum of the $\bfu_c$ modes. Because $\bfu_c$ modes are not passively tracing $\bfu_s$ and are the only modes that cause $\bnabla\rho$, any mass density-related spectrum is going to be mostly sensitive to the $\bfu_c$ mode spectrum, hence we expect that dust continuum and column density spectra should be closer to $\ekin(k) \sim k^{-2}$, which they are for the Large Magellanic Cloud (LMC) \citep{Colman2022_large_scale_driving}, and too in the post-shock regions in the Cygnus Loop \citep{Raymond2020_cygnus_loop_shock}. 

    \subsection{IK spectrum: scale-dependent alignment and energy flux}
        IK-type spectra have been motivated through a number of different phenomenological and analytical means  \citep{Iroshnikov_1965_IK_turb,Kraichnan1965_IKturb,Hosking2020_tangled_field_stats,Galtier2023_IK_spectrum}, but given that Figure~\ref{fig:spectra}~(e) shows that on $k > \ks$ the turbulence is dominantly incompressible and $\ekin(k) \sim \emag(k)$ in amplitude, the simplest explanation is Alfv\'enic turbulence where shear Alfv\'en modes are dynamically aligned into Alfv\'enic $\bfu\propto\pm\bfb$ states, i.e., dynamical alignment \citep{Boldyrev2006}. Dynamical alignment predicts $\emag(k)\sim\ekin(k) \sim \varepsilon^{2/3}k^{-5/3}\theta_{\bfu,\bfb}^{-2/3}$ for purely Alfv\'enic MHD turbulence, where $\theta_{\bfu,\bfb}$, the angle between $\bfu$ and $\bfb$, follows a scale-dependent relation $\theta_{\bfu,\bfb}(\ell) \sim \ell^{1/4}$ (defined in Appendix~\ref{app:scale_dependent_dfn}) which results in $\emag(k)\sim \ekin(k) \sim \varepsilon^{2/3}k^{-3/2}$. Hence, alignment makes the Alfv\'enic cascade less efficient, creating a shallower $k^{-3/2}$ spectrum. If we adopt the relation $\ekin(k) \sim \varepsilon^{2/3}k^{-5/3}\theta_{\bfu,\bfb}^{-2/3}$, we demonstrate below that we obtain results that are consistent with aspects of our numerical findings. To test this, we calculate the probability distribution function of $\theta_{\bfu,\bfb}$, $p(\theta_{\bfu,\bfb})$ and $|\theta_{\bfu,\bfb}(\ell)|$ structure function and show them in panels (a) and (b) in Figure~\ref{fig:alignment_between_ub}, respectively. 
        
        In (a) we show a strong preference for $\bfu\propto\pm\bfb$ Alfv\'enic states, and then in (b) we see further that the $\bfu$ and $\bfb$ become preferentially aligned on small $\ell$, following a $\theta_{\bfu,\bfb}(\ell) \sim \ell^{1/8}$ dependence, only on $\ell < \ell_s$ corresponding to the $\ekin(k) \sim k^{-3/2}$ and $\emag(k) \sim k^{-9/5}$ cascades. Combining $\theta_{\bfu,\bfb}(\ell)\sim\ell^{1/8}$ with a $\ekin(k) \sim \varepsilon^{2/3}k^{-5/3}\theta_{\bfu,\bfb}^{-2/3}$ spectra, i.e., assuming that all the velocity motions are Alfv\'enic, $\ekin(k) \sim \varepsilon^{2/3}k^{-19/12}$, inconsistent with our data. One potential way of resolving this discrepancy with this specific spectral model is to assume that the kinetic energy fluxes are scale-dependent, $\varepsilon \sim \ell^{\beta}$ -- a significant departure from standard turbulence phenomenology -- such that $\ekin(k) \sim (k^{-\beta})^{2/3}k^{-5/3}(k^{-1/8})^{-2/3}$. This can be motivated by considering that the magnetic dynamo may tap into the kinetic energy reservoir in a scale-dependent manner, depleting the $\varepsilon$ fluxes differently on each $k$ mode. For $\beta = -1/8$, we get a $\ekin(k) \sim k^{-3/2}$ spectrum, as desired. We show the kinetic, $\Pi_{uu}(k)$ and magnetic $\Pi_{bb}(k)$ energy flux functions in panels (c) and (d) in Figure~\ref{fig:spectra}, respectively. $\Pi_{bb}(k)$ is approximately constant across the $\emag(k) \sim k^{-9/5}$ spectrum, whilst $\Pi_{uu}(k)$ is scale-dependent, following $\Pi_{uu}(k) \sim k^{1/8}$, as required for the $\ekin(k) \sim k^{-3/2}$ spectrum based on the $\ekin(k) \sim \varepsilon^{2/3}k^{-5/3}\theta_{\bfu,\bfb}^{-2/3}$ model. 
        
\section{Discussion}\label{sec:discussion}
        By running supersonic MHD turbulence simulations at unprecedented grid resolutions of up to $10,\!080^3$, approaching realistic Reynolds numbers for the cold ISM, and larger Reynolds numbers than in the warmer phases \citep{Ferriere2020_reynolds_numbers_for_ism}, we have revealed the existence of two scale-separated $\ekin(k)$ cascades: (1) a Burgers-type spectrum ($\sim k^{-2}$), which hosts $\M > 1$ kinetic-energy-dominated motions $\ekin(k) > \emag(k)$, which non-locally transports energy to all scales, similar to supersonic hydrodynamical turbulence \citep{Galtier2011_exact_relations,Ferrand2020_10k3_fluxes}; and (2) an IK-type spectrum ($\sim k^{-3/2}$), which hosts $\M < 1$ magnetically-dominated, $\ekin(k) < \emag(k)$ mostly incompressible motions $\nabla\cdot\bfu \approx 0$, that undergo a local cascade to smaller scales and progressively become more aligned with $\bfb$, $\theta_{\bfu,\bfb}(\ell) \sim \ell^{1/8}$. The energy flux over the whole kinetic energy cascade is scale-dependent, $\Pi_{uu}(k) \sim k^{1/8}$.  The first clear indication that MHD energy fluxes may feature scale-dependent properties was recently reported in \citep{Grete2023_as_a_matter_of_dynamical_range} using the same energy flux transfer functions that we use. Weak scale-dependent energy flux has been found previously in supersonic, hydrodynamic turbulence \citep{Ferrand2020_10k3_fluxes}.  
             
        Moreover, when $\theta_{\bfu,\bfb}(\ell) \sim \ell^{1/8}$ and $\Pi_{uu}(k) \sim k^{1/8}$ are combined together, this significant violation from textbook turbulence turns a $\sim k^{-5/3}$ spectrum into an IK spectrum. The relation between $\Pi_{uu}(k)$ and $\theta_{\bfu,\bfb}$ is directly related to the kinetic energy reservoir being depleted in a scale-dependent manner through $\bfu$ and $\bfb$ alignment turning off $\bnabla\times(\bfu\times\bfb)$ -- the magnetic flux generated by the dynamo. It is suppressed the strongest on small scales, $\ell < \ell_{\rm eq}$, where $\bfu$ and $\bfb$ are the most parallel, and the weakest at large scales $\ell \approx \ell_{\rm eq}$, where the magnetic flux is being replenished. We observe no evidence of spectral steepening from current sheet instabilities \citep{Dong2022_reconnection_mediated_cascade,Galishnikova2022_saturation_and_tearing}, potentially due in part to the weak $\theta_{\bfu,\bfb}(\ell) \sim \ell^{1/8}$ alignment, compared to $\sim \ell^{1/4}$, which changes the instability criterion for the anisotropic turbulent eddies \citep{Loureiro2017_reconnection_in_turbulence,Comisso2018_MHD_turbulence_plasmoid_regime}. Hence, the hierarchy of scales in this turbulence regime is, $\ell_0 \approx \ell_{\rm cor} > \ell_{\rm super} >\ell_{\rm eq} > \ell_{\rm s} > \ell_{\rm sub} > \ell_{\nu}$, where $\ell_{\rm super}$ are the scales for the $\M > 1$ cascade, where in real space $\delta u(\ell_{\rm super}) \sim \ell_{\rm super}^{1/2}$, and $\ell_{\rm sub}$ are the scales for the $\M < 1$ cascade, where in real space $\delta u(\ell_{\rm sub}) \sim \ell_{\rm sub}^{1/4}$. 
    
        We find consistent results (within 1$\sigma$) for the position of $\ell_{\rm s}$ with previous hydrodynamical simulations \citep{Federrath2021}, but with a shallower slope in the subsonic cascade ($k^{-3/2}$ compared to $k^{-5/3}$), meaning that more kinetic energy is available on these scales compared to hydrodynamical turbulence. In the cold ISM, $\ks$ is roughly the filament width scale \citep{Arzoumanian2011,Federrath2016_filaments,Andre2022_filament_widths}, hence on scales smaller than a typical filament width, the turbulence becomes highly magnetized, which in turn prevents small-scale cloud fragmentation via the additional magnetic pressure, and because of the shallower $\sim k^{-3/2}$ kinetic energy spectrum, the small scales sequester more turbulent support, suppressing the star formation process \citep{Padoan2011,Federrath2012}. In the volume-filling warm ionized phase (WIM) $\ks$ is roughly on the outer scale \citep{Gaesnsler_2011_trans_ISM}, hence the strong magnetic fields that we see grown via the turbulent motions are efficient enough to maintain a strong, energy equipartition magnetic field through the whole medium, with a local, forward ($\varepsilon > 0$) $\sim k^{-3/2}$ cascade in kinetic energy and $\emag(k) \sim k^{-9/5}$ in magnetic energy, aligned in a scale-dependent manner $\theta_{\bfu,\bfb}\sim\ell^{1/8}$.
        
        For $\emag(k)$, we find a single, local, forward cascade with a $\emag(k) \sim k^{-9/5}$ spectrum. This is significantly different from the $\ekin(k)$ spectrum, with the power law only emerging on $\M < 1$ scales, necessitating a separate theoretical treatment for the two cascades \citep{Grete2021_as_a_matter_of_tension}, even on the scales where $\ekin(k)\sim\emag(k)$. The hierarchy of scales for the turbulent magnetic field is then $\ell_0 > \ell_{\rm cor} \approx \ell_{\rm eq} > \ell_{\rm s} > \ell_{\rm cascade} > \ell_{\eta}$, corresponding to real space $\delta b(\ell_{\rm cascade}) \sim \ell_{\rm cascade}^{2/5}$. This is steeper than the classical Alfv\'enic theories \citep{Goldreich1995,Boldyrev2006}, and shallower than the tearing instability theories \citep{Dong2022_reconnection_mediated_cascade,Galishnikova2022_saturation_and_tearing}, making the $\emag(k)$ generated from a turbulent dynamo potentially unique. Indeed, this spectrum is consistent within the $1\sigma$ uncertainties for the turbulent magnetic field spectra derived from rotation measure structure functions observed in the ISM for the Small Magellanic Cloud $\emag(k) \sim k^{-1.3\pm0.4}$, and very close to those found in the LMC $\emag(k) \sim k^{-1.6\pm0.1}$ \citep{Seta2023_structure_function}, possibly indicating that the magnetic field in these satellite galaxies is being maintained by a turbulent dynamo as in our simulation.

        With the two cascades in the kinetic energy, scale-dependent energy flux, $\bfu$ and $\bfb$ alignment, and the single cascade in the magnetic energy, this study presents a new paradigm for compressible turbulence in the ISM, with a magnetic field maintained by a dynamo. We hope not only that these results stimulate further fundamental, theoretical investigations, which are required to derive the $\emag(k) \sim k^{-9/5}$ and $\theta_{\bfu,\bfb}\sim\ell^{1/8}$ relations, but that there is an effort to directly determine if this spectrum is consistent with rotation measure structure function observations from ongoing observational campaigns and next-generation radio telescopes, like ASKAP's Polarisation Sky Survey of the Universe's Magnetism (POSSUM) \citep{Anderson2021_possum_survey,Vanderwoude2024_ASKAP_possum} and the SKA Observatory, which would provide direct evidence that the magnetic field of the ISM of our Galaxy is maintained by the chaotic, turbulent motions from within it. 

\backmatter

\bmhead{Acknowledgments}
    We acknowledge the useful discussions with Drummond Fielding, Alexander Chernoglazov and Andrey Beresnyak on the local anisotropy and alignment structure functions, and the more general discussions about this work with Riddhi Bandyopadhyay, Philipp~K.-S.~Kempski, Eliot Quataert, Alexander Philippov, Philip Mocz, Bart Ripperda and Chris Thompson. \textbf{Funding:} J.~R.~B.~acknowledges financial support from the Australian National University, via the Deakin PhD and Dean's Higher Degree Research (theoretical physics) Scholarships and the Australian Government via the Australian Government Research Training Program Fee-Offset Scholarship and the Australian Capital Territory Government funded Fulbright scholarship. J.~R.~B., C.~F., R.~S.~K. and S.~C. further acknowledge high-performance computing resources provided by the Leibniz Rechenzentrum and the Gauss Centre for Supercomputing grants~pr32lo, pr73fi and GCS large-scale project~10391.
    C.F.~acknowledges funding by the Australian Research Council (Discovery Projects grants~DP230102280 and DP250101526), and the Australia-Germany Joint Research Cooperation Scheme (UA-DAAD). C.F.~further acknowledges high-performance computing resources provided by the Australian National Computational Infrastructure (grant~ek9) and the Pawsey Supercomputing Centre (project~pawsey0810) in the framework of the National Computational Merit Allocation Scheme and the ANU Merit Allocation Scheme.
    R.~S.~K.~acknowledges support from the European Research Council via the ERC Synergy Grant “ECOGAL” (project ID 855130), from the German Excellence Strategy via the Heidelberg Cluster of Excellence (EXC 2181 - 390900948) “STRUCTURES”, and from the German Ministry for Economic Affairs and Climate Action in project “MAINN” (funding ID 50OO2206). R.~S.~K.~also thanks for local computing resources provided by the Ministry of Science, Research and the Arts (MWK) of {\em The L\"{a}nd} through bwHPC and the German Science Foundation (DFG) through grant INST 35/1134-1 FUGG and 35/1597-1 FUGG, and also for data storage at SDS@hd funded through grants INST 35/1314-1 FUGG and INST 35/1503-1 FUGG. J.~.R.~B. and A.~B further acknowledge the support from NSF Award 2206756. 
    \textbf{Author Contributions:} J.~R.~B.~led the entirety of the project, including the GCS large-scale project~10391, running the simulations, co-developing the \textsc{FLASH} code and analysis programs used in this study, and led the writing and ideas presented in the manuscript. C.~F.~co-led the GCS large-scale project~10391, is the lead developer of the \textsc{FLASH} code and the analysis pipelines used in the study, and contributed to the ideas presented in this study and drafting of the manuscript. R.~S.~K.~co-led the GCS large-scale project~10391, and contributed to the ideas presented in this study and drafting of the manuscript. S.~C.~provided invaluable technical advice and assistance during the GCS large-scale project proposal and during the run time of the simulations, provided support visualizing the large datasets, and contributed to the ideas presented in this study and drafting of the manuscript. A.~B.~contributed to the ideas presented in this study and drafting of the manuscript. \textbf{Competing interests:} We declare no competing interests. \textbf{Data and materials availability:} All raw data for temporally-averaged energy spectra, probability distributions functions and structure functions presented in the study are available at the \textsc{GitHub} repository: https://github.com/AstroJames/10k\_supersonicMHD. \textbf{License information:} Copyright ©2024 the authors, some rights reserved. 

\newpage
\begin{appendices}
\input{appendix}
\end{appendices}

\bibliography{sn-bibliography}

\end{document}

%% file: header.tex
\newcommand{\bfu}{\bm{u}}

\newcommand{\bfr}{\bm{r}}
\newcommand{\bfell}{\bm{\ell}}
\newcommand{\bfk}{\bm{k}}
\newcommand{\bfb}{\bm{b}}
\newcommand{\bfj}{\bm{j}}

\newcommand{\bnabla}{\bm{\nabla}}
\newcommand{\cs}{c_s}
\newcommand{\M}{\mathcal{M}}
\newcommand{\Ma}{\mathcal{M}_{\rm A}}
\newcommand{\ekin}{\mathcal{E}_{\rm kin}}
\newcommand{\emag}{\mathcal{E}_{\rm mag}}

\newcommand{\kperp}{k_{\perp}}
\newcommand{\Rm}{\mathrm{Rm}}
\renewcommand{\Re}{\mathrm{Re}}

\newcommand{\Exp}[2]{\left\langle{#1}\right\rangle_{#2}}
\renewcommand{\d}[1]{\ensuremath{\operatorname{d}\!{#1}}}
\newcommand{\dthree}[1]{\ensuremath{\operatorname{d}^3\!{#1}}}

\newcommand{\keq}{k_{\rm eq}}
\newcommand{\kcor}{k_{\rm cor}}
\newcommand{\ks}{k_{\rm s}}
\newcommand{\ku}{k_{\rm u}}

%% file: appendix.tex
\section{Online Methods}\label{app:simulations}
\subsection{Basic numerical model and code:} 
    We use a modified version of the magnetohydrodynamical (MHD) code \textsc{flash} \citep{Fryxell2000,Dubey2008}. Our code uses a highly-optimized, hybrid-precision \citep{Federrath2021}, positivity-preserving, second-order MUSCL-Hancock HLL5R Riemann scheme \citep{Bouchut2010,Waagan2011} to solve the compressible, ideal, MHD fluid equations in three dimensions,
    \begin{align}
    \partial_t \rho + \nabla\cdot\left(\rho \bfu\right) = 0,& \label{eq:continuity}\\
    \partial_t\!\left(\rho \bfu\right) + \nabla\cdot\left(\rho\bfu\!\otimes\!\bfu + p\mathbb{I} - \frac{1}{\mu_0}\bfb\!\otimes\!\bfb \right) = \rho \bm{ f},&\label{eq:momentum} \\
    \partial_t \bfb  + \nabla\cdot(\bfu\otimes\bfb - \bfb\otimes\bfu) = 0,& \label{eq:induction}\\ 
    \nabla\cdot\bfb = 0,\;\; p = \cs^2\rho+ \frac{1}{2\mu_0}\bfb \cdot \bfb,& \label{eq:pressure}
    \end{align}
    where $\rho$, $\bfu$, $\bfb$ and $\mu_0$ are the mass density, the velocity and magnetic fields, and the magnetic permittivity, respectively. Equation~\ref{eq:pressure} relates the scalar pressure $p$ to $\rho$ via the isothermal equation of state with constant sound speed $c_s$, as well as the pressure contribution from the magnetic field. We work in units $c_s = \rho_0 = \mu_0 = L = 1$, where $\rho_0$ is the mean mass density and $L$ is the characteristic length scale of the system, such that $L^3 = \mathcal{V} = 1$ is the volume. We discretize the equations over a triply-periodic domain of $[-L/2, L/2]$ in each dimension, with grid resolutions $2,\!520^3$, $5,\!040^3$ and $10,\!080^3$ -- the largest grids in the world for simulations of this fluid turbulence regime. In order to drive turbulence, a turbulent forcing term $\bm{f}$ is applied in the momentum equation (details below). This set of equations including the forcing term is the standard approach in modeling driven, magnetized turbulence. These calculations were only possible as part of a large-scale high performance computing project, large-scale project~10391, at the Leibniz Supercomputing Centre in Garching, Germany. They were run on the supercomputer SuperMUC-NG. For the $10,\!080^3$ simulation, and the power-spectrum calculations, we utilized close to $140,\!000$ compute cores, and close to 80~million compute-core hours. 

\subsection{Turbulent driving:} 
    We choose to drive the turbulence with a turbulent Mach number of $\M \approx 4$ to ensure that we resolve a sufficient range of both supersonic $\delta u > c_s$ and subsonic $\delta u < c_s$ scales \citep{Federrath2021}. We apply a non-helical stochastic forcing term $\bm{f}$ in Equation~\ref{eq:momentum}, following an Ornstein-Uhlenbeck stochastic process \citep{Eswaran1988_forcing_numerical_scheme,Schmidt2009,Federrath2010_solendoidal_versus_compressive}, using the \textsc{TurbGen} turbulence driving module \citep{Federrath2010_solendoidal_versus_compressive,Federrath2022_turbulence_driving_module}. The forcing is constructed in Fourier space such that kinetic energy is injected at the smallest wavenumbers, peaking at $L\ell_0^{-1} = k_0L/2\pi = 2$ and tending to zero parabolically in the interval $1\leq kL/2\pi\leq3$, allowing for self-consistent development of turbulence on smaller scales, $kL/2\pi>3$, as routinely performed in turbulence box studies. To replenish the large-scale compressible modes and shocks, we decompose $\bm{f}$ into its incompressible ($\nabla\cdot\bm{f}=0$) and compressible ($|\nabla\times\bm{f}|=0$) mode components \citep{Federrath2010_solendoidal_versus_compressive}, and drive the turbulence with equal amounts of energy in each of the modes, termed ``mixed" or ``natural" driving \citep{Federrath2008}. Note that even though we drive with a ``natural'' mix of modes, the velocity modes, even at low-$k$ do not perfectly match the energy distribution of the driving modes. See \citet{Federrath2010_solendoidal_versus_compressive} for the exact nonlinear relation between the driving modes and the velocity modes.  

\subsection{Initial conditions and hierarchical interpolation:}\label{app:init_conditions}
    We initialize $\rho(x,y,z) = \rho_0$ and $\bfu = \bm{0}$. The total magnetic field $\bfb = \bfb_0 + \bfb_{\rm turb}$ is composed of both a mean (external or guide) field $\bfb_0$ and turbulent $\bfb_{\rm turb}$ component. The $\bfb_{\rm turb}$ evolves self-consistently with the MHD turbulence via Equation~\ref{eq:induction}. For our simulations, $\bfb_0 = 0$, and only the isotropic, turbulent magnetic field remains, $\bfb = \bfb_{\rm turb}$. Regardless of the initial field amplitude, the same small-scale dynamo saturation is reached \citep{Beattie2023_growth_or_decay}. The same also holds for different seed magnetic fields \citep{Seta2020_seed_magnetic_field}. Hence, given enough integration time, we can initialize a magnetic field with any initial structure and amplitude and be confident that it will result in the same saturation. In order to limit the use of computational resources on the fast or nonlinear dynamo stages \citep{Federrath2016_dynamo,Rincon2019_dynamo_theories}, we initialize the magnetic field amplitude and structure close to the saturated state for the $2,\!520^3$ simulation. From our previous experiments at lower resolutions, this is $\emag/\ekin \approx 1/4$ (or $\Ma \approx 2$, where $\Ma = \delta u / \delta v_{\rm A}$ is the Alfv\'en Mach number), and with a significant amount of power at all $k$.
    
    Driven MHD turbulence in this regime takes $\approx (1-2)t_0$, where $t_0 = \ell/\delta u$ is the turbulent turnover time on the outer scale, to shed the influence of its initial conditions and establish a stationary state \citep{Federrath2010_solendoidal_versus_compressive,Price2010_grid_versus_SPH,Beattie2022_spdf}. To avoid expending compute resources on simulating this transient state, we only apply the previously discussed initial conditions to the $2,\!520^3$ simulation. For the remaining $5,\!040^3$ and $10,\!080^3$ simulations, we interpolate the initial conditions hierarchically from the simulations with lower resolutions, i.e., we initialize the $5,\!040^3$ simulation with linearly interpolated initial conditions from the $t \approx 2t_0$ state of the $2,\!520^3$ simulation and the $t \approx 3t_0$ state of the $5,\!040^3$ simulation for the $10,\!080^3$ simulation. We use linear interpolation to preserve $\nabla\cdot\bfb = 0$ between grid interpolations. It takes a tiny fraction of $t_0$, $t \sim \Re^{-1/2}t_0$, to populate the new modes after the interpolation onto the higher-resolution grid. Hence this provides an adequate method for minimizing the amount of compute time spent making the $5,\!040^3$ and $10,\!080^3$ simulations stationary.

\subsection{Estimating the Reynolds numbers:}\label{app:reynolds_numbers}
    Our numerical model is an implicit large eddy simulation (ILES), which relies upon the spatial discretisation to supply the numerical viscosity and resistivity as a fluid closure model. Recently, a detailed characterization of our code's numerical viscous and resistive properties, specifically for turbulent boxes, has been performed by comparing the ILES model with direct numerical simulations (DNS), which have explicit viscous and resistive operators in Equation~\ref{eq:momentum} and Equation~\ref{eq:induction}, respectively \citep{Kriel2022_kinematic_dynamo_scales,Grete2023_as_a_matter_of_dynamical_range,Shivakumar2023_numerical_dissipation}. \citet{Shivakumar2023_numerical_dissipation} derived empirical models for transforming grid resolution $N_{\rm grid}$ into $\Re$ and $\Rm$. For supersonic MHD turbulence, they find $\Re = (N_{\rm grid}/N_{\Re})^{p_{\Re}}$, where $p_{\Re} \in [1.5, 2.0]$ and $N_{\Re} \in [0.8, 4.4]$ and $\Rm = (N_{\rm grid}/N_{\Rm})^{p_{\Rm}}$, where $p_{\Rm} \in [1.2, 1.6]$ and $N_{\Rm} \in [0.1, 0.7]$. For our three resolutions, $N_{\rm grid}=2,\!520$, $5,\!040$ and $10,\!080$, this gives $\Re_{2 \rm k} \in [1.8 \times 10^5, 3.3 \times 10^5]$, $\Re_{5 \rm k} \in [5\times 10^5 , 1.3 \times 10^6]$, $\Re_{10k} \in [1.4\times 10^6, 5.3\times 10^6]$ and $\Rm_{2\rm k} \in [1.9\times10^5, 4.9\times10^5]$, $\Rm_{5 \rm k} \in [4.4\times10^5, 1.5\times10^6]$, $\Rm_{10\rm k} \in [1\times10^6, 4.5\times10^6]$, where the $2\rm k$, $5\rm k$ and $10\rm k$ subscripts correspond to the $N_{\rm grid} = 2,\!520$, $5,\!040$ and $10,\!080$ simulations, respectively. When we report the Reynolds numbers in the main text, we report the average between the bounds placed on each of the dimensionless plasma numbers.
    
\subsection{Data structure and domain decomposition:} 
    \textsc{flash} uses a block-structured parallelization. Each 3D computational block is distributed onto one single compute core. For the $10,\!080^3$ simulation, we use $168\!\times\!210\!\times\!210$ cells per block, resulting in $(168\!\times\!60, 210\!\times\!48, 210\!\times\!48) = (10080, 10080, 10080)$ cells in each spatial direction, for a total of $60\!\times\!48\!\times\!48=138,\!240$~cores in the $10,\!080^3$ run. The $5,\!040^3$ and $2,\!520^3$ simulations, which serve to check numerical convergence of statistical quantities, have a block structure $(84\!\times\!60, 210\!\times\!24, 210\!\times\!24) = (5040, 5040, 5040)$, using $34,\!560$~cores, and $(42\!\times\!60, 210\!\times\!12, 210\!\times\!12) = (2520 , 2520, 2520)$, using $8,\!640$~cores, respectively.

\subsection{File I/O:} 
    \textsc{flash} is parallelised with \textsc{mpi}. File I/O is based on the \textsc{hdf}\oldstylenums{5} library. Since our runs will use $138,\!240$~cores (3024~compute nodes on SuperMUC-NG) and produce approximately $100$ output files with $29\,\mathrm{TB}$ each (approximately $3.0\,\mathrm{PB}$ in total), efficient file I/O is extremely important. In order to achieve the highest efficiency when reading and writing these huge files, we use parallel-\textsc{hdf}\oldstylenums{5} together with a split-file approach. In this approach each core writes simultaneously to disk, grouping data from 288~cores together into a total of 504~files per output dump. This proved to be an extremely efficient method, providing an I/O throughput that is close to the physical maximum of approximately $200\,\mathrm{GB/s}$ reachable on the SuperMUC-NG \textsc{/scratch} file system. The net effect is that I/O only takes about 3--4~minutes to read or write a $28\,\mathrm{TB}$ checkpoint file, such that it consumes only a minor fraction of the resources compared to the integration of the MHD fluid equations.

\subsection{Major code optimizations:} 
    Our version of \textsc{flash} is highly optimized for solving large-scale hydrodynamical and MHD problems \citep{Federrath2021}. Specifically, the number of stored 3D fields are reduced to the bare minimum required for these simulations (only the mass density and three velocity and magnetic field components are stored). All calls to the equation of state routines are performed inline, directly in the Riemann solver. The code is precision hybridized such that all fluid variables are stored in single precision (4~bytes per floating-point number), but critical operations are performed in double-precision arithmetic (8~bytes per floating-point number), which retains the accuracy of the full double-precision computations. These efforts significantly reduce the computational time and the required amount of \textsc{mpi} communication, as well as the overall memory consumption. In addition, the single-precision operations benefit from a higher SIMD count and lower cache occupancy, for a further parallel speedup. As a result the code is almost $4\times$ faster and requires $4.1\times$ less memory than the \textsc{flash} public release, while retaining the full accuracy. Previous studies have further characterized the performance and comparison of our code with the public \textsc{flash} version \citep{Cielo2020_code_scaling}.

    \begin{figure}
        \centering
        \includegraphics[width=\linewidth]{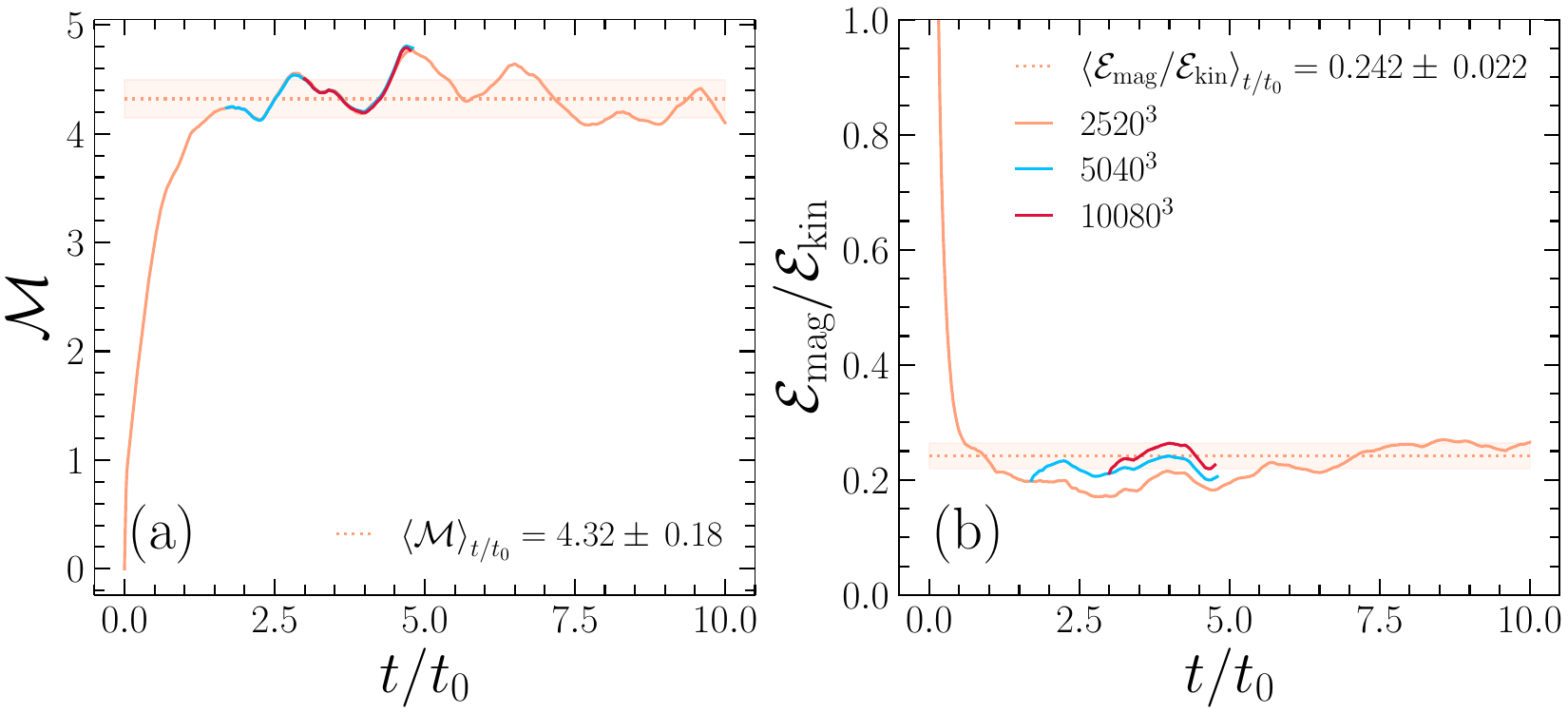}
        \caption{\textbf{The time evolution of the volume integral root-mean-squared velocity fluctuations and magnetic to kinetic energy ratios.} \textbf{(a):} The turbulent Mach number $\M$ as a function of time in units of correlation times $t_0$ for the $2,\!520^3$ (orange), $5,\!040^3$ (blue) and $10,\!080^3$ (red) simulations. Each higher-resolution simulation is carried out from an initial condition of a linearly interpolated version of the lower-resolution simulation, avoiding the initial non-stationary state, i.e., the $0 \leq t/t_0 \lesssim 2$ range of times for the $2,\!520^3$ simulation. The average and 1$\sigma$ are shown for the last $5t_0$ with the dashed line and bounding box, respectively, giving $\M = 4.32 \pm 0.18$. \textbf{(b):} the same as (a) but for the integral magnetic to kinetic energy ratio, $\emag/\ekin$. In the steady state, $\emag/\ekin = 0.242 \pm 0.022$, as desired (see \ref{app:init_conditions}).}
        \label{fig:integral_quantities}
    \end{figure}

\section{Time evolution of integral quantities}\label{app:volume_integral_quants}
        We show the time evolution of $\M$ (panel a) and $\emag/\ekin$ (panel b) in and Figure~\ref{fig:integral_quantities}, showing the time-averaged value for each of the quantities in the legend. The different colors represent the different simulation grids and, as discussed in the previous section, from this plot it is easy to observe where in time each of the simulations were started from, using the hierarchical interpolation technique discussed in the previous section. These plots convey that all the simulations are indeed in a stationary state. Between the different grids, the values of $\M = 4.32 \pm 0.18 \approx 4$, shown in panel (a) are nearly perfectly matching over time, likely due to the $\M \propto (\int\d{k}\;\ekin(k))^{1/2}$ being dominated by the low-$k$ velocity modes, so changing the grid spacing, has little effect on the $\M$ statistics. However, there are minor differences in $\emag/\ekin = 0.242 \pm 0.022 \approx 1/4$, shown in panel (b) as we change the simulation resolution, with $\emag/\ekin$ growing as we increase the resolution. This can be attributed to the fact that the magnetic field, and hence $\emag(k)$, is inherently a small-scale, high-$k$ mode-dominated field, $\kcor \approx 10$, in comparison to the velocity field $\kcor \approx 2$. This feature is well-recognized in the dynamo community \citep{Schekochihin2004_dynamo,Federrath2016_dynamo,Galishnikova2022_saturation_and_tearing,Beattie2023_growth_or_decay}, and we show it explicitly in Figure~\ref{fig:spectra} where we separate the $\emag(k)$ and $\ekin(k)$ spectra. Note that $\emag/\ekin = 0.242 \pm 0.022 \approx 1/4$ is roughly an order of magnitude higher than the $\emag/\ekin$ value for $\M \approx 4$ turbulence driven with a natural mix in previous studies \citep{Federrath2011_mach_dynamo}. This is because our simulation resolves a huge range of highly-magnetized scales (see Figure~\ref{fig:spectra}), scales smaller than the energy equipartition and the sonic scale. This drives $\emag/\ekin$ up such that the Alfv\'enic Mach number is $\Ma \approx 2 \iff \emag/\ekin \approx 1/4$.

    \begin{figure}
        \centering
        \includegraphics[width=0.8\linewidth]{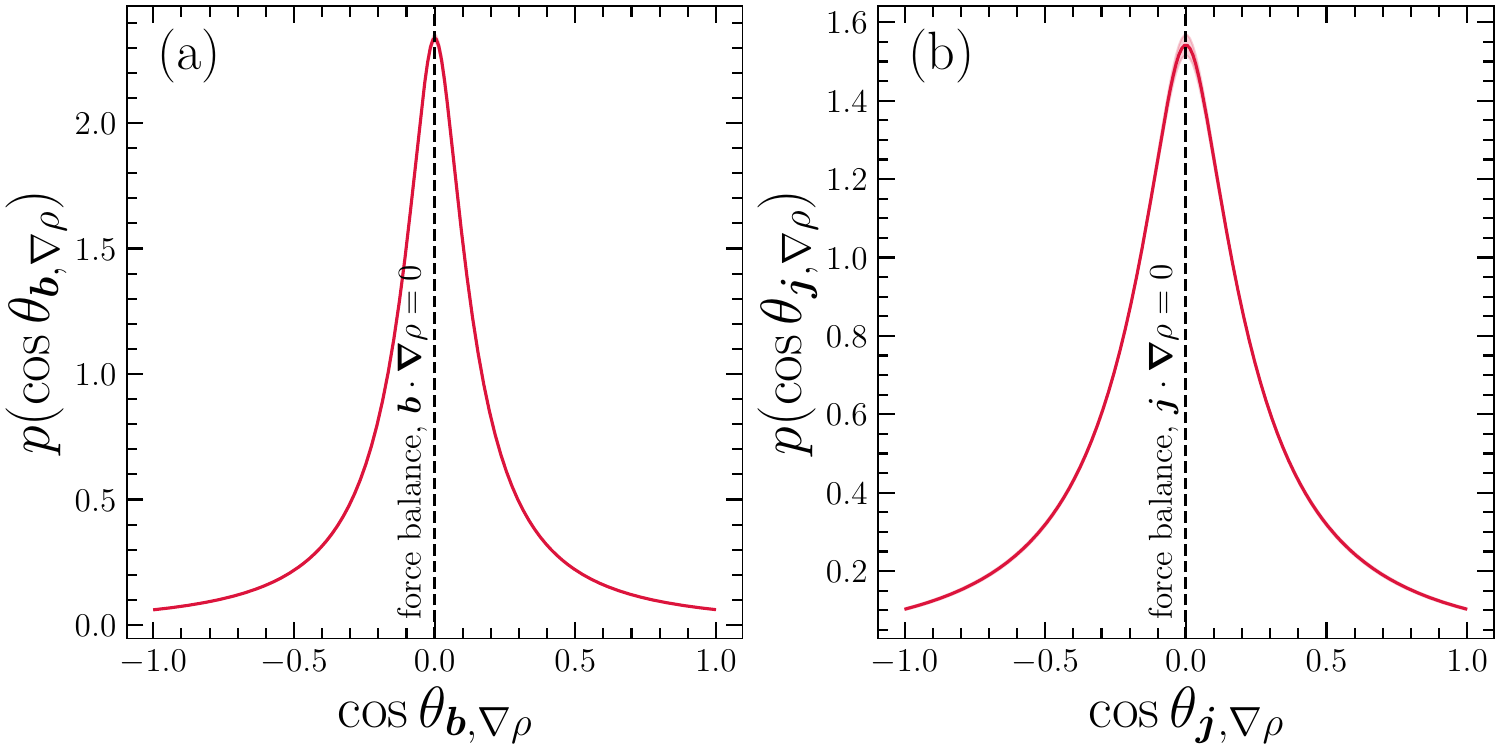}
        \caption{\textbf{Alignment between the current density and gradient of the mass density, showing that the turbulence is on average in force balance, which fixes the current sheets to mass density gradients.} The probability distribution function for the cosine angle between the (a) magnetic field, (b) current density and the gradient of the mass density for the $10,\!080^3$ supersonic MHD simulation. The vertical dashed line shows that the turbulence favors a state in force balance, $\bfj \times \bfb \propto \bm{\nabla} \rho$, where both $\bfb$ and $\bfj$ are orthogonal to $\bm{\nabla}\rho$, in turn, making the average current sheet sensitive to the effects of compressibility.}
        \label{fig:force_balance_pdf}
    \end{figure}

\section{Force-balanced current structures and instabilities}\label{app:current_structure}         
        The current sheet structure shown in Figure~\ref{fig:slice_poster} is correlated with the mass density structure. In local force balance $\bm{\nabla} \rho \propto \bfj \times \bfb$, which can happen in a fraction of a sound crossing of a shocked region \citep{Robertson2018,Mocz2018}. We show this is the average behavior of the turbulence by plotting the probability distribution function of the alignment between $\bfj$ and $\bm{\nabla}\rho$ in Figure~\ref{fig:force_balance_pdf}, revealing a strong peak at $\bfb\cdot\bm{\nabla} \rho = \bfj\cdot\bm{\nabla} \rho = 0$. Thus, it is possible that the supersonic motions that drive large $|\bm{\nabla}\rho|$ significantly disrupt the small-scale current sheets in the plasma, confining them to regions where there is large $|\bm{\nabla}\rho|$. The particular configuration between $\bm{\nabla}\rho$ and $\bfb$ has been observed before in Planck polarization observations of the dense, molecular ISM \citep{Planck2016a,Soler2017}, and force-balanced sheets permeating through the ISM have been assumed in recent ISM scintillation models \citep{Kempski2024_propagation_and_scattering}. Hence, complex networks of intense sheets of current are on average fixed along the mass density filaments, correlating the effects of compressibility with current sheets. 

\begin{figure}
        \centering
        \includegraphics[width=\linewidth]{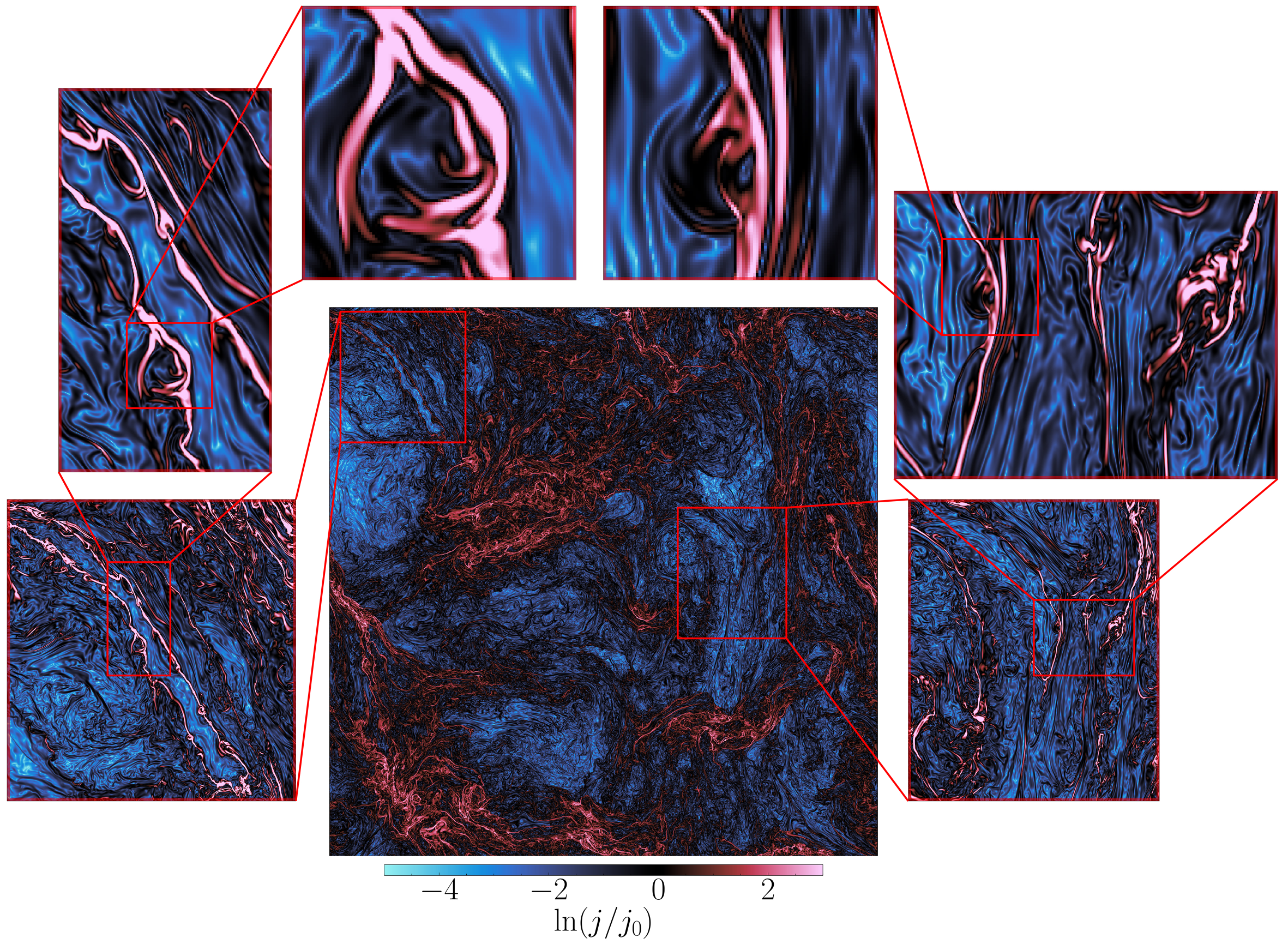}
        \caption{\textbf{Tearing instabilities forming in current sheets in the voids of magnetized ISM turbulence.} A two-dimensional logarithmic current density slice using the same color scheme as shown in the top of Figure~\ref{fig:slice_poster}. Three levels of zoom-ins reveal chains of tearing unstable modes appearing in thin current sheets situated in the mass-density voids. We zoom directly into one of the unstable modes appearing in each chain, with the maximum zoom-in revealing the entire cross-section of the mode. The size scale of this region is $k_{*}L/2\pi\approx100$. The well-structured chains appear in the most isolated sheets, away from the large, intense current structure that forms alongside the dense filament, suggesting that disruptive events from the shocks prevent the formation of tearing-unstable sheets everywhere in the supersonic plasma.}
        \label{fig:plasmoids}
    \end{figure}

        Further inspection of $\ln(j/j_0)$ in Figure~\ref{fig:slice_poster} reveals that there are chaotic current structures generated in the dense shocked regions and low-volume-filling, more linearly-structured sheets developing tearing instabilities sparsely throughout the mass density voids (colored blue), where the shear flow is smallest. The lifetime of the voids can be orders of magnitude longer than that of the shocked regions \citep{Hopkins2013_non_lognormal_s_pdf,Robertson2018,Mocz2019,Beattie2022_spdf}, making voids excellent environments for the development of current sheet instabilities. We show a zoom-in of specific unstable current sheets that have developed the classical chain-like structure, usually indicative of the plasmoid instability in Figure~\ref{fig:plasmoids} \citep{Bhattacharjee2009_fast_reconnection,Loureiro_2016_plasmoid_instability,Dong2018_role_of_plasmoid_instability}. We use the outer scale of these unstable modes to estimate $\ell_{*}^{-1} \sim k_{*}L/2\pi \gtrsim 10^2$, which are well-resolved in our simulations. In the supersonic regime, if these instabilities are constrained to sparsely populate the voids, this may prevent them from influencing the global volume-weighted statistics of the turbulence, as in \citep{Dong2022_reconnection_mediated_cascade, Galishnikova2022_saturation_and_tearing}, but makes them viable candidates for the low-volume-filling intermittent structures required for strong cosmic-ray scattering in the ISM \citep{Kempski2022_cosmic_ray_scattering,Fielding2022_ISM_plasmoids,Kempski2023_b_field_reversals,Butsky2024_cr_scattering_with_intermittency}. 

\section{Definition of energy spectra and turbulent scales}\label{app:turbulent_scales}
    \subsection{Energy spectra:}
    The magnetic energy spectrum is defined as 
    \begin{align}
        \emag(k) = \frac{1}{2\mu_0}\int\d{\Omega_k}\;  \Tilde{\bfb}(\bm{k})\Tilde{\bfb}^{\dagger}(\bm{k}) 4\pi k^2,
    \end{align}
    and kinetic energy spectrum is defined as,
    \begin{align}
        \ekin(k) = \frac{\rho_0}{2}\int \d{\Omega_k}\; \Tilde{\bfu}(\bm{k}) \Tilde{\bfu}^{\dagger}(\bm{k}) 4\pi k^2,
    \end{align}
    where the tilde indicates the Fourier transform of the underlying field variable, dagger the complex conjugate and the $\int \d{\Omega_k}4\pi k^2$ is the shell integral over fixed $k$ shells, producing isotropic, one-dimensional energy spectra. Note that other definitions of the kinetic energy spectrum have been used in the literature, which define $\bm{w} = \sqrt{\rho}\bfu$ and then take the square Fourier transform of $\bm{w}$ to construct the kinetic energy spectrum \citep{Federrath2010_solendoidal_versus_compressive,Kritsuk2007,Federrath2013_universality,Grete2017_shell_models_for_CMHD,Grete2020_as_a_matter_of_state,Grete2023_as_a_matter_of_dynamical_range}. However, we pick the simplest definition of $\ekin(k)$ to allow us to more easily compare with theories of incompressible turbulence \citep{Iroshnikov_1965_IK_turb,Kraichnan1965_IKturb,Goldreich1995,Boldyrev2006}, and even compressible theories \citep{Bhattacharjee_1998_weakly_compressible_solar_wind,Lithwick2001_compressibleMHD}, which adopt the same $\ekin(k)$ definition as we do in this study.

    \subsection{Inner and outer scales:}
         We use the following definitions for the inner and outer scales of the turbulent cascades, using $u$ and $b$ superscripts to differentiate between the kinetic energy and magnetic energy scales, respectively. For both the $\emag(k)$ and $\ekin(k)$, the outer scale is directly related to the integral or correlation scale of the energy spectrum, 
        \begin{align}
            \frac{\kcor L}{2\pi} = \frac{\displaystyle \int\d{k}\,\mathcal{E}(k)}{\displaystyle \int\d{k}\, (kL/2\pi)^{-1}\mathcal{E}(k)},  
        \end{align}
        which for $\ekin(k)$ closely tracks the driving scale $\kcor^u =  (2.03 \pm 0.01)2\pi/L \approx k_0L2/\pi = 2$, but for $\emag(k)$ depends on a range of parameters, like the growth stage of the dynamo and the strength of the large-scale magnetic field \citep{Beattie2022energy_balance,Beattie2023_growth_or_decay}. For our $10,\!080^3$ simulation, $\kcor$ of the magnetic field is $\kcor^b = (12.05 \pm 0.05)2\pi/L$, highlighting how the magnetic field is intrinsically a small-scale field. For the inner scale we take the maximum of the $(kL/2\pi)^2\mathcal{E}(k)$ spectrum,
        \begin{align}
            \frac{\ku L}{2\pi} = \left(\frac{\max\left\{(kL/2\pi)^2\mathcal{E}(k)\d{k}\right\}}{\displaystyle \int\d{k}\,\mathcal{E}(k)}\right)^{1/2},
        \end{align}
        which probes the smallest scale of the magnetic and velocity gradients, since, e.g.\ for the velocity, $k^2\ekin(k)\d{k} \sim k^2 u^2 \sim (\nabla\otimes\bfu)^2$, defining the end of the turbulent cascade and the start of diffusion-dominated scales. We show both of these scales in panel (a) and (c) in Figure~\ref{fig:spectra}. For our $10,\!080^3$, the velocity inner scale is $\ku^u L/2\pi = 525\pm12$, and the magnetic inner scale is $\ku^b L/2\pi=399\pm10$.
        
    \begin{figure}
        \centering
        \includegraphics[width=\linewidth]{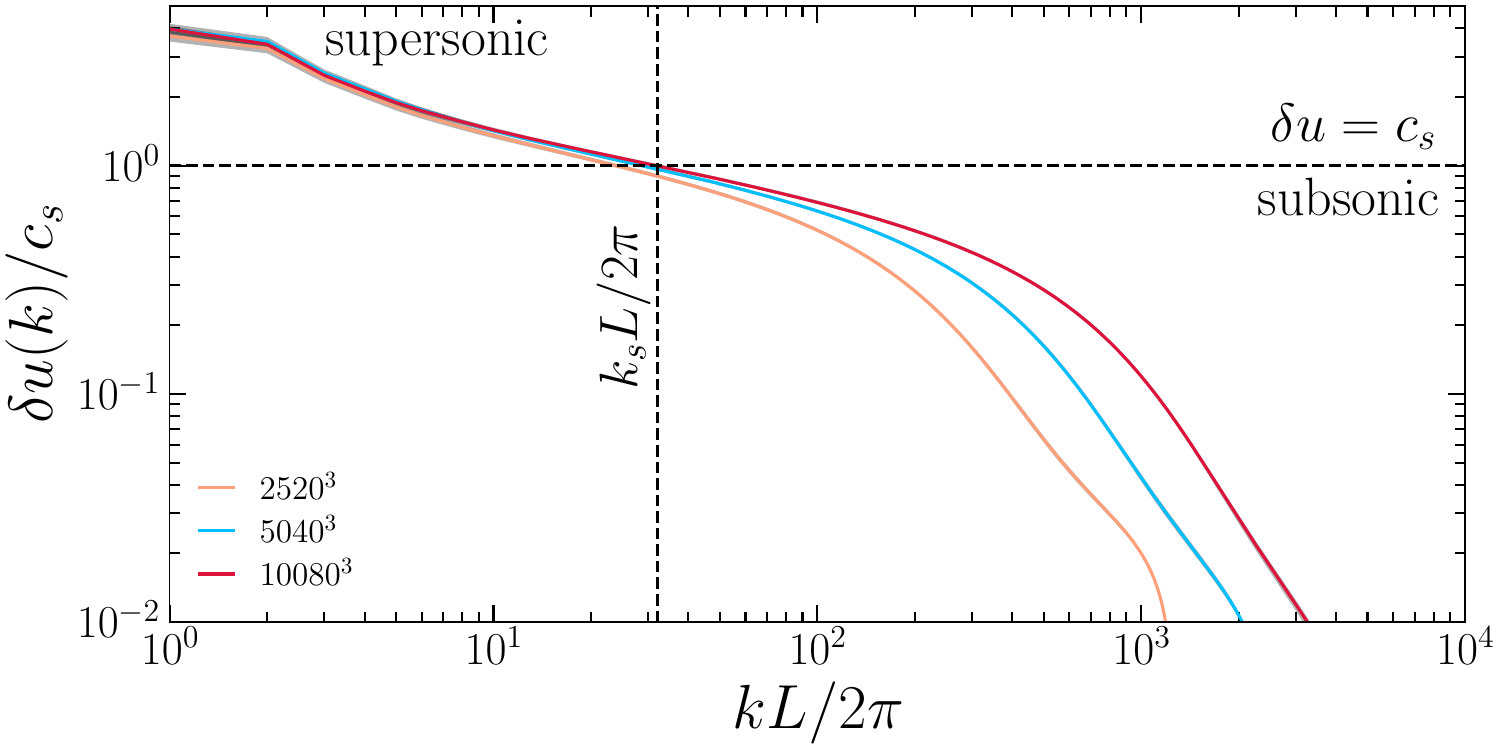}
        \caption{\textbf{Velocity dispersion as a function of wavenumber and the sonic scale.} The root mean square (rms) velocity normalized by the sound speed, $\delta u /c_s$, as a function of wavenumber $k$. The wavenumber where $\delta u = c_s$, the sonic scale $\ks L/2\pi$, is indicated with a horizontal, black, dashed line, with $\delta u > c_s$, corresponding to supersonic rms velocities, and $\delta u < c_s$ to subsonic rms velocities. The sonic scale, $\ks L/2\pi = 32 \pm 2$ for the $10,\!080^{3}$ simulation, is indicated with a vertical dashed line.}
        \label{fig:k_space_vel_dispersion}
    \end{figure}

    \subsection{The sonic scale:} 
        The first transition that we investigate is the $k$-space sonic scale, denoted as $\ks L/2\pi$. Using Parseval's theorem, we calculate the rms velocity as a function of scale through the relation
        \begin{align}\label{eq:k_space_vel}
        \frac{\delta u(k)}{c_s} = \left(\frac{2}{\rho_0 }\int^{\infty}_{k}\d{k'}\,\ekin(k')\right)^{1/2} = \M(k),
        \end{align}
        where $\ks L/2\pi$ is the $k$ mode where $\delta u(k_{\rm s}) = c_s$ \citep{Federrath2010_solendoidal_versus_compressive,Federrath2012,Federrath2021}. We present this calculation in Figure~\ref{fig:k_space_vel_dispersion} and determine the $\ks L / 2\pi$ root for the sonic transition
        \begin{align}\label{eq:sonic_scale}
            \delta u(k)/c_s - 1 = 0,
        \end{align}
        which is indicated by the dashed line. For the $10,\!080^3$ simulation, we find $\ks L/2\pi = 32 \pm 2$, for $5,\!040^3$, $\ks L/2\pi = 29 \pm 2$ and $2,\!520^3$, $\ks L/2\pi = 24 \pm 2$. For $k>\ks$, the plasma becomes subsonic with $\delta u(k)/c_s < 1$, and for $k<\ks$, the plasma is supersonic with $\delta u(k)/c_s > 1$. This is marked by the horizontal black dashed line. As previously found, there is a smooth transition between these two flow regimes rather than a sharp discontinuity \citep{Federrath2021}. 
    
    \subsection{The energy equipartition (or MHD) scale:}
        The second transition that we study is the transition from kinetic energy dominated turbulence, $\ekin(k) > \emag(k)$ to magnetic energy dominated turbulence, $\emag(k) > \ekin(k)$, which is equivalent to comparing the turbulent Alfv\'en timescale, $t_{\rm A} = \ell/\delta v_{\rm A}(\ell)$, where $\delta v_{\rm A}$ is the rms Alfv\'en velocity of the plasma, with the turbulent velocity timescale $t_{\rm turb} = \ell/\delta u(\ell)$. If $\ekin(k) > \emag(k)$, then $t_{\rm turb} < t_{\rm A}$ and vice versa for $\ekin(k) < \emag(k)$. For strong guide field turbulence, this has been previously called the MHD scale \citep{Goldreich1995,Lithwick2001_compressibleMHD}, but we use the more general energy equipartition scale nomenclature, $\keq L/2\pi$, as in the main text. Even though comparing these timescales looks like a calculation about critical balance, since both timescales are describing intrinsically nonlinear fluctuations, this is not a probe for weak versus strong turbulence \citep{Perez2008_weak_and_strong_turb,Meyrand2016_strong_to_weak_transition}. To determine $\keq L /2\pi$ we find the root of $\emag(k)/\ekin(k) - 1 = 0$. We plot the full $\emag(k)/\ekin(k)$ spectrum and $\keq L /2\pi$ mode in panel (d) of Figure~\ref{fig:spectra}. For the $10,\!080^3$ simulation, we find $\keq L/2\pi = 10.6 \pm 0.7$, for $5,\!040^3$, $\keq L/2\pi = 11.9 \pm 0.7$ and $2,\!520^3$, $\keq L/2\pi = 13.1 \pm 0.6$. 

\section{Energy Flux Transfer Functions}\label{app:transfer_functions}
    Energy transfer functions are invaluable statistics that probe the nature of local energy flux between modes. They have been applied to turbulent dynamo and driven turbulence in the past, \cite{Mininni2005_transfer_functions,Alexakis2005_shell_to_shell}, and were recently generalized for fully compressible fluid plasmas \cite{Grete2017_shell_models_for_CMHD}. In the second row of Figure~\ref{fig:spectra} we plot the energy flux transfer functions, probing the 3-mode transfer of energy flux. We follow the definitions for compressible magnetohydrodynamics \cite{Grete2017_shell_models_for_CMHD}. For both the advective and compressive kinetic energy transfers, they are given,
    \begin{align}
    &\quad\quad\quad\quad\quad\quad\bm{w}'' \xrightarrow{\bfu} \bm{w}' \nonumber \\
        \mathcal{T}_{uu}(k',k'') =& -\int\dthree{\bfell}\; \left\{\bm{w}'\otimes\bfu:\bnabla\otimes\bm{w}'' + \frac{1}{2}\bm{w}'\cdot \bm{w}'' (\bnabla\cdot\bfu)\right\},
    \end{align}
    where $\bm{w} = \sqrt{\rho}\bfu$, and the fields $\bm{w}'\equiv \bm{w}(\bfk')$ and $\bm{w}'' \equiv \bm{w}(\bfk'')$ are the fields filtered over those modes. The filter is defined isotropically,
    \begin{align}
        \bm{w}'(\bfell) = \int\dthree{\bfk}\;\delta(|\bfk| - |\bfk'|) \Tilde{\bm{w}}(\bfk) \exp\left\{ 2\pi i \bfk\cdot\bm{\ell} \right\},
    \end{align}
    where, $\bfk'$ is a shell in $k$ space (i.e., a collection of modes). The shells are defined logarithmically, 
    \begin{align}
        \left\{ |\bfk'| \right\} = \left\{ |\bfk''| \right\} &= \left\{ 2^{ (i -1) / 4 + 2} \right\},\quad\quad i = 0,\; \dots,\; 4 \frac{\ln(N_{\rm grid}/8)}{\ln(2)} + 1,
    \end{align}
    which results in the best localization of eddy-type interactions \cite{Grete2017_shell_models_for_CMHD}. We likewise define the transfer of energy flux for the magnetic energy,
    \begin{align}
        & \quad\quad\quad\quad\quad\quad\bm{b}'' \xrightarrow{\bfu} \bm{b}' \nonumber \\
        \mathcal{T}_{bb}(k',k'') =& -\int\dthree{\bfell}\;\left\{ \bfb'\otimes\bfu:\bnabla\otimes\bfb'' + \frac{1}{2}\bfb'\cdot \bfb'' (\bnabla\cdot\bfu)\right\}.
    \end{align}
    In panels (c) and (d) of Figure~\ref{fig:spectra}, we use these flux transfer functions to directly identify (1) the nature of locality of the kinetic and magnetic cascade and (2) whether or not a cascade exists, corresponding to neighboring $k'$ and $k''$ shells receiving and donating energy flux, respectively. Still following \citep{Grete2017_shell_models_for_CMHD}, we define the cross-scale energy flux as, 
    \begin{align}
        \Pi_{XX}(k) = \sum_{k < k''}\sum_{k \geq k'}\mathcal{T}_{XX}(k',k''),
    \end{align}
    which is constant across $k$ modes if the energy flux between modes is constant, as is assumed in all turbulence theories \citep{Schekochihin2020_bias_review}. 
    
\section{Empirical measurements for the slopes of the energy spectra}\label{app:slopes}
    In Figure~\ref{fig:spectra} and the corresponding section, we provide tilde slopes, accompanied by compensations in each of the panels. Here we directly report the slopes utilizing weighted linear least squares on the linearized counterpart of the model $\mathcal{E}(k) = \beta_0 k^{\beta_1}$. For the weights, we use the 1$\sigma$ from the time-averaged spectra. For the kinetic energy spectra we partition the $k$ domain into supersonic scales within the supersonic cascade, $4 \leq k_{\rm super}L/2\pi \leq \keq L/2\pi$, and subsonic scales, within the subsonic cascade $\ks L/2\pi = 33 \leq k_{\rm sub}L/2\pi \leq 10 \ks L/2\pi = 330$. We find, in general, our choices for the fit domain do not have a large impact on the exact values, as long as we pick scales within the cascades. For $\ekin(k_{\rm super})$ we find $\beta_1 = -2.01 \pm 0.03$, and for $\ekin(k_{\rm sub})$, $\beta_1 = -1.465 \pm 0.002$, close to the tilde values we present in the main text, $\sim -2$ and $\sim -3/2$, respectively. Performing the same analysis for the compressible $\ekin^{\rm comp}$ and solenoidal $\ekin^{\rm sol}$ mode kinetic energy spectra we find for $\ekin^{\rm comp}(k_{\rm super})$ $\beta_1 = -2.05 \pm 0.04$, for $\ekin^{\rm comp}(k_{\rm sub})$, $\beta_1 = -1.971 \pm 0.001$, $\ekin^{\rm sol}(k_{\rm super})$ $\beta_1 = -1.97 \pm 0.05$ and $\ekin^{\rm sol}(k_{\rm sub})$ $\beta_1 = -1.425 \pm 0.001$, reinforcing that the compressible modes follow a single spectrum $\sim k^{-2}$, not passively tracing the incompressible modes, and the incompressible modes capture the supersonic-to-subsonic dichotomy. We do the same fits to the magnetic spectra over the single domain $80 \leq  k L/2\pi \leq 250$, and find $\beta_1 = -1.798 \pm 0.001$, consistent with the $\sim k^{9/5}$ scaling we compensate the spectra by in the main text.

\section{Definition of alignment structure function}\label{app:scale_dependent_dfn}
    To compute the alignment structure function shown in Figure~\ref{fig:alignment_between_ub} (b), we first define our increments,
    \begin{align}
        \bm{\delta}\bfu &= \bfu(\bfr) - \bfu(\bfr + \bfell), \\
        \bm{\delta}\bfb &= \bfb(\bfr) - \bfb(\bfr+\bfell),
    \end{align}  
    for separation vector $\bm{\ell}$. Next, we define a local mean magnetic field direction,
    \begin{align}
        \widehat{\bfb}_{\bfell} = \frac{\bfb(\bfr) + \bfb(\bfr + \bfell)}{\|\bfb(\bfr) + \bfb(\bfr + \bfell)\|},
    \end{align}
    and then find the perpendicular component to the local field for each of the fluid variables, e.g., for $\bfu$ and $\bfb$,
    \begin{align}
        \bm{\delta}\bfu_{\perp} &= \bm{\delta}\bfu - (\bm{\delta}\bfu\cdot \widehat{\bfb}_{\bfell})\widehat{\bfb}_{\bfell}, \\
        \bm{\delta}\bfb_{\perp} &= \bm{\delta}\bfb - (\bm{\delta}\bfb\cdot\widehat{\bfb}_{\bfell})\widehat{\bfb}_{\bfell},
    \end{align}
    which is the standard definition for these quantities \citep{Chernoglazov2021_alignment_SR_MHD,Dong2022_reconnection_mediated_cascade}. Next we construct the ratio between first-order structure functions,
    \begin{align}
    |\theta_{\bfu,\bfb}(\ell)| \sim |\sin\theta_{\bfu,\bfb}(\ell)| = \frac{\Exp{|\bm{\delta}\bfu_{\perp} \times \bm{\delta}\bfb_{\perp}|}{\ell}}{\Exp{ | \bm{\delta}\bfu_{\perp} | | \bm{\delta}\bfb_{\perp} |}{\ell}},
    \end{align}
    We use $2 \times 10^{12}$ sampling pairs to ensure that the structure functions are converged at all scales \citep{Federrath2021}. Furthermore, as with the spectra, we construct the structure functions across a number of realizations in the stationary state, and then time-average the structure function to produce Figure~\ref{fig:alignment_between_ub}. 